# Simultaneous 3D co-registered perfusion and oxygenation with ULM, photoacoustic imaging, and a planar matrix array

**Léa Davenet[1,2], Jacques Battaglia[1], Franck Lager[3], Pascal Dargent[1], Charlotte Lussey-Lepoutre[2,4], Bertrand Tavitian[5,6], Olivier Couture[1], S. Lori Bridal[1] and Jérôme Gateau[1,*]**

[1] Sorbonne Université, CNRS, Inserm, Laboratoire d'Imagerie Biomédicale, LIB, F-75006, Paris, France
[2] Sorbonne Université, Université Paris Cité, Inserm, Centre de Recherche des Cordeliers, Equipe Labellisée Ligue contre le Cancer, F-75006 Paris, France
[3] Université Paris Cité, CNRS, Inserm, Institut Cochin, F-75014 Paris, France
[4] Sorbonne Université, Department of Nuclear Medicine, Hôpital Pitié Salpêtrière, APHP, 75013 Paris France
[5] Université Paris Cité, Inserm, PARCC, F-75015 Paris, France
[6] Service de Radiologie, AP-HP, Hôpital européen Georges Pompidou, F-75015 Paris, France
*Author to whom any correspondence should be addressed.

**E-mail:** jerome.gateau@cnrs.fr

## Abstract

*Objective.* Joint assessment of tissue oxygenation and microvascular perfusion could offer valuable insights into vascular function across a wide range of biomedical applications. Multispectral photoacoustic imaging enables the evaluation of blood oxygenation, while ultrasound localization microscopy provides sub-diffraction visualization of the microvasculature and blood perfusion. Here, we combine these two complementary modalities to simultaneously generate co-registered, volumetric maps of blood oxygenation and perfusion.

*Approach.* Photoacoustic imaging and ultrasound localization microscopy are both ultrasound-based techniques. We developed an imaging platform that integrates the two modalities using a single planar ultrasonic matrix array, a state-of-the-art array for 3D ultrasound localization microscopy. The bimodal platform was validated *in vitro* using vessel-mimicking phantoms, *then in vivo* in mice.

*Main results. In vitro* bimodal images of tubes injected with contrast agents demonstrated a co-registration accuracy of 20 µm and revealed complementary structural and functional information. Multispectral photoacoustic imaging achieved oxygen saturation measurements spanning the physiological range (60–95 %) with 5 % accuracy using only five optical wavelengths. *In vivo* imaging of healthy mouse tissues with known vascular anatomy further demonstrated the ability of the proposed platform to jointly characterize blood oxygenation and microvascular perfusion.

*Significance.* This work experimentally validates a bimodal photoacoustic imaging–ultrasound localization microscopy approach using a planar ultrasound array. We characterized the functional imaging performance of this platform and identified limited-view artifacts inherent to this array configuration in photoacoustic imaging. These findings establish a foundation for adopting the platform in future studies of murine models and for advancing this promising bimodal approach.

## 1. Introduction

Tissue oxygenation and microvascular perfusion are tightly coupled under normal physiological conditions: a reduction in oxygen availability or an increase in local metabolic demand typically triggers an increase in microvascular blood flow [1]. However, this tight coupling between oxygen delivery and metabolic demand can be disrupted in pathologies with microcirculatory dysfunction or metabolic dysregulation. For instance, in solid tumours, vascular structure abnormalities (such as microvascular shunts) may induce heterogeneous perfusion and local hypoxia [2]. Detection and characterization of these dysfunctions, to select appropriate therapeutic strategies, require the assessment of both tissue oxygenation and microvascular perfusion.

Several imaging modalities allow sequential or concurrent evaluation of perfusion and oxygenation. Multiparametric magnetic resonance imaging (MRI) is particularly versatile. Dynamic contrast-enhanced MRI provides measurements of tissue perfusion, blood volume fraction, and capillary permeability [3], while blood oxygenation level-dependent imaging enables indirect assessment of tissue oxygenation [4]. While MRI enables deep-tissue imaging, its resolution is on the millimetre scale, and quantitative oxygenation assessment is both challenging and model-dependent. Positron emission tomography (PET) can provide absolute quantification of perfusion ($H_2{}^{15}O$ PET) [5], oxygen extraction fraction, metabolic rate of oxygen ($^{15}O_2$ PET) [6], and hypoxia [7]. Despite its excellent depth capability, PET suffers from limited spatial resolution (≈4–6 mm), involves ionizing radiation, and may require the injection of multiple radiotracers to obtain both perfusion and oxygenation maps. Lastly, optical methods such as diffuse optical tomography (DOT) using near-infrared (NIR) spectroscopy [8] map tissue oxygenation by estimating oxyhaemoglobin ($HbO_2$) and deoxyhaemoglobin (Hb) concentrations. Temporal fluctuations in total haemoglobin may serve as indirect perfusion indicators, although DOT is limited to superficial tissues (≤2–3 cm) and provides only moderate spatial resolution (≈5–10 mm). Thereby, mapping both tissue oxygenation and microvascular perfusion with high spatial resolution at depths of several centimetres and using a single imaging modality remains highly challenging.

A promising high-resolution approach combines photoacoustic (PA) imaging, which assesses tissue oxygenation, with ultrasound localization microscopy (ULM), which maps microvascular perfusion. ULM is a subwavelength ultrasound (US) localization technique. It works by localizing and tracking intravascular microbubbles across sequential pulse-echo US images. This allows vascular networks to be reconstructed beyond the US diffraction limit, revealing microvascular structure and flow [9,10]. Initially developed in two dimensions, ULM has since been extended to 3D [11], enabling whole-organ microvascular mapping at resolutions down to a few tens of micrometres [12,13]. This resolution is sufficient to resolve individual renal glomeruli in both animals and humans [14]. A similar level of detail has been reached in the brain, down to individual capillaries [15]. Beyond structure, functional ULM reveals brain-wide neurovascular activity at microscopic scale [16]. Together, these advances are now paving the way for clinical translation [17].

Photoacoustic imaging (PAI), or optoacoustic imaging, complements ULM by enabling non-invasive quantification of blood oxygenation with submillimetre resolution at depths of several centimetres [18,19]. PAI is based on the photoacoustic effect: pulsed optical excitation of haemoglobin generates ultrasonic waves with amplitudes proportional to the local optical absorption coefficient. Using multispectral PAI, which involves successive acquisitions at different optical wavelengths, $HbO_2$ and Hb can be spectrally distinguished to quantify relative concentrations and to directly estimate blood oxygen saturation ($SO_2$). Therefore, combining ULM and PAI in a single acquisition sequence with a single transducer array offers unique advantages: intrinsic co-registration of microvascular structures and oxygenation maps as well as direct analysis of perfusion-oxygenation coupling at the microvascular scale.

Although both state-of-the-art ULM and multispectral PAI can provide volumetric images using matrix arrays, the ideal detection configuration for both modalities (a large number of sensitive transducers distributed over a wide angular aperture with fine spatial sampling and driven in parallel) is not yet feasible. This limitation has led to the employment of modality-specific strategies best adapted to each technique. Volumetric ULM commonly uses planar matrix arrays and high-frame-rate imaging with multiple ultrasound wavefront emissions to maximize the contrast-to-noise ratio and ensure robust detection of microbubbles behaving as isolated scatterers [20]. Due to the directional emission and relatively low amplitude of PA waves generated from elongated structures, PAI systems employing matrix arrays generally rely on spherical or hemispherical detection geometries with large, highly sensitive transducers to mitigate limited-view artifacts [21–24]. These conflicting strategies lead to trade-offs for the development of 3D combined ULM and PAI systems.

Recent advances in 3D PA-ULM implementations have adopted state-of-the-art PAI spherical detection geometries with the main goal of murine brain imaging through intact scalp and skull. The first demonstration of co-registered volumetric images combining microbubble-enhanced power Doppler, ULM, and PAI at 800 nm using a spherical array was reported in 2022 [25]. In this study, ULM imaging was achieved using US emissions from a subset of array transducers to reach a lower frame rate (50 Hz) than that obtained with multiplexed matrix arrays [11,26,27]. Using a similar approach in a murine model of ischemic stroke, Tang *et al.* [24] were able to obtain a higher US frame rate of 215 Hz, allowing quantification of cerebral blood perfusion, vascular density, and blood flow velocity via ULM alongside $SO_2$ mapping with PAI at three optical wavelengths. This revealed stroke-induced ischemia, hypoxia, and flow reduction. The same platform was subsequently applied to investigate glymphatic function [28]. Using a NIR-II-absorbing PA dye, cerebrospinal fluid circulation was imaged through the intact skull, revealing impaired glymphatic function after ischemic stroke, aging, and anaesthesia. Alternative approaches relied on external ultrasound sources, such as laser-generated ultrasound [29] or single-element transducers [30], enabling frame rates of 100 Hz while relying on a single transmit wavefront. While these studies demonstrate the feasibility of 3D ULM and PAI *in vivo* and provide high-quality images, they sacrifice US image quality for ULM because the matrix array is primarily optimized for PAI.

In this study, we developed and characterized a 3D PA-ULM implementation based on a state-of-the-art ULM detection geometry: a high-density planar matrix array. In contrast to previous implementations, which were primarily optimized for PAI, our approach prioritizes ULM performance at the expense of PA image quality. Fully sampled planar matrix arrays were previously used *in vivo* to combine PA volumetric imaging with US imaging [31,32], although the US images were restricted to conventional B-mode anatomical imaging rather than ULM. Other approaches have relied on mechanically scanned conventional linear US arrays to perform sequential cross-sectional acquisitions combining ULM and multispectral PAI. These methods have recently been validated for exploring the mouse kidney [33] and brain hemodynamic responses during oxygen challenges [34]. These studies leverage US technology that enabled the development of ULM [35] and demonstrate the strong potential of combining ULM and PAI. However, stacking cross-sectional imaging introduces artifacts in ULM due to out-of-plane motion [36,37] and loss of microbubble correspondence across imaging planes, resulting from limited elevational resolution in 2D acquisitions. In parallel, limited-view artifacts in PAI [38] further prevent volumetric reconstruction.

Here, we present a novel integrated approach for simultaneous ULM and multispectral PAI using a single fully sampled planar US matrix array, previously validated for 3D ULM. This single-array configuration provides intrinsically co-registered datasets that combine high-resolution microvascular perfusion mapping with functional blood oxygenation assessment. The contributions of this work are: (i) the first demonstration of simultaneous 3D multispectral PA and ULM imaging with a single planar US matrix array to assess microvascular perfusion and blood oxygenation; (ii) the generation of intrinsically co-registered volumetric datasets, enabling direct voxel-wise comparison of structural and functional information; and (iii) experimental validation on tissue-mimicking phantoms as well as mouse tails and hindlimbs *in vivo*, demonstrating complementary, high-resolution microvascular function characterization.

## 2. Materials and methods

### 2.1. Photoacoustic evaluation of whole blood oxygenation

Within the NIR range, haemoglobin is the primary endogenous optical absorber in biological tissues at depths beyond the skin barrier. Moreover, the optical absorption of whole blood is primarily due to haemoglobin molecules. In blood, two forms of haemoglobin coexist, in varying proportions depending on the measurement site: $HbO_2$ and Hb. The two forms exhibit different optical absorption spectra. Therefore, the absorption spectrum of blood depends directly on its oxygen saturation level ($SO_2$), which is defined as the ratio of $HbO_2$ to total haemoglobin and is expressed as:

$$SO_2 = \frac{c^{HbO_2}}{c^{tot}} = \frac{c^{HbO_2}}{c^{HbO_2} + c^{Hb}} \quad (1)$$

Where $c^{HbO_2}$, $c^{Hb}$ and $c^{tot}$ denote the molar concentrations of $HbO_2$, Hb, and total haemoglobin, respectively. At an optical wavelength λ and an oxygen saturation level $SO_2 \in [0,1]$, the optical absorption coefficient of whole blood $\mu_a^{wb}$ can be written as (2):

$$\mu_a^{wb}(\lambda, SO_2) = c^{tot} \cdot \left[SO_2 \cdot \epsilon_a^{HbO_2}(\lambda) + (1 - SO_2) \cdot \epsilon_a^{Hb}(\lambda)\right] \cdot \ln(10) \tag{2}$$

where $\epsilon_a^{HbO_2}(\lambda)$ and $\epsilon_a^{Hb}(\lambda)$ are the molar extinction coefficients of $HbO_2$ and Hb, respectively.

This spectral difference permits $SO_2$ quantification using optical techniques that require at least two optical wavelengths. For PAI, we have recently determined [39] that the relevant wavelength range for $SO_2$ measurement spans 700-850 nm. Moreover, we extracted the five most commonly used wavelengths for the spectral separation of $HbO_2$ and Hb, namely $\lambda_1$ = 700 nm, $\lambda_2$ = 730 nm, $\lambda_3$ = 760 nm, $\lambda_4$ = 800 nm, and $\lambda_5$ = 850 nm. These wavelengths were used here for *in vivo* spectral PA acquisitions.

To validate our PA method *in vitro* within the $SO_2$ range most frequently observed in clinical practice, we performed estimations of $SO_2$ on blood-mimicking solutions reproducing $SO_2$ levels between 60 % and 100 %. Indeed, arterial oxygen saturation in healthy individuals typically ranges from 95 % to 100 % [40]. Mixed venous oxygen saturation ($SvO_2$) measures oxygen saturation in blood entering the pulmonary artery before oxygenation in the lungs and reflects the global balance between systemic oxygen delivery and consumption. Under normal physiological conditions, $SvO_2$ is approximately 70-75 % [41]. When oxygen delivery becomes insufficient relative to metabolic demand, tissues compensate by increasing the extraction of oxygen from haemoglobin. As the organism is capable of extracting 50-60 % of the delivered oxygen [41], $SvO_2$ may fall to 40-50 % in pathological conditions.

**2.2. Experimental imaging setup**

Figure 1(a, b) illustrates the experimental setup and the imaging phantom in the configuration used for the *in vitro* study (described in Section 2.4). Both PAI and ULM were performed using a single US system: a 1 024-element matrix probe (centre frequency: 7.8 MHz; -6 dB bandwidth: 60 %; Vermon, Tours, France) connected to a Vantage research scanner equipped with 256 elements (Verasonics, Kirkland, WA, USA) via a Verasonics UTA 1 024 MUX adapter. The matrix elements were divided into four panels, each containing 256 elements, with discontinuities between panels along the y-axis (Figure 1(c)). The transducer elements were arranged on a 32 × 35 grid (dimensions: x × y) with a 300 µm pitch. Each element measured 0.3 × 0.3 $mm^2$. Columns 9, 17, and 25 along the y-axis were left empty, yielding a total field of view of 9.6 × 10.5 $mm^2$. The resulting configuration provided a full acceptance angle of approximately 35° (at -6 dB) in both elevation and lateral directions at the centre frequency. The origin of the coordinate system (cross in Figure 1(c)) was set at the centre of the probe, with the plane z = 0 corresponding to the probe surface.

For PAI, the setup included two additional components: 1) an optical excitation system, and 2) a synchronization interface. Optical excitation was provided by a tuneable optical parametric oscillator laser (SpitLight 600 OPO, Innolas Laser GmbH, Krailling, Germany; wavelength range 680-980 nm, <8 ns pulse duration, 20 Hz pulse repetition frequency). Laser light was delivered through a bifurcated fibre bundle (CeramOptec GmbH, Bonn, Germany). Two protected silver mirrors (PFSQ05-03-P01, Thorlabs Inc., Newton, NJ, USA) deflected the output beams to provide bilateral illumination of the imaged volume on opposite sides of the US probe. Each arm of the fibre bundle terminated in a compact 3D-printed mount, ensuring fixed positions and orientations of the fibre tips and mirrors. The laser beams intersected the matrix probe axis 10 mm below the probe surface. A mechanical assembly integrating the US probe and fibre mounts ensured consistent alignment between the US window and the illumination. To prevent undesired illumination of the probe surface and associated parasitic signals, the US window was covered with a metalized Mylar film (space blanket), which is transparent to US in the relevant frequency range and opaque to laser light. Laser pulse energy was monitored using the pyrometer integrated into the laser system, used as an internal energy meter. An optical fluence of approximately 7 mJ $cm^{-2}$ was measured at 730 nm in air at the intersection point of the two illumination beams. Synchronization between optical excitation and US acquisition was managed by an external trigger generator (BNC Model 577, Berkeley Nucleonics, San Rafael, CA, USA). This generator ensured synchronization between the laser's Pockels cell and the US scanner following each firing of the laser flash lamp. The laser pulse repetition frequency defined the time base of the PAI sequence.

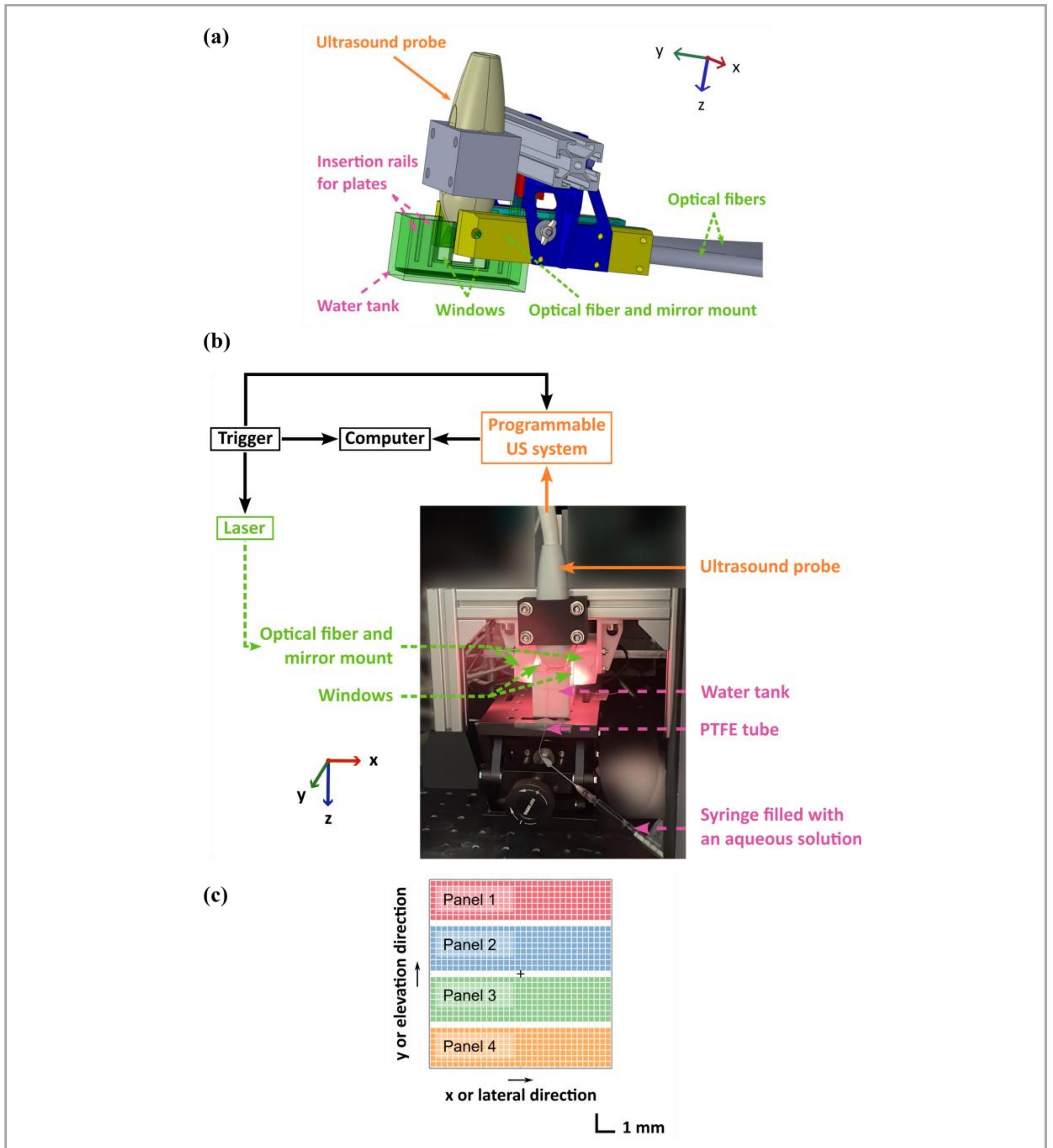

**Figure 1.** Experimental setup with the micro-vessel imaging phantom. *(a)* Schematic of the mechanical support securing the positions of the 7.8 MHz ultrasound matrix probe and of the light delivery system (optical fibers and mirrors in the yellow holder). Optical excitation (700-850 nm) is applied bilaterally. The water tank (green, with side windows) contains the imaging phantom and is held in place by the mechanical assembly. *(b)* Photograph of the experimental setup. Blood vessels are mimicked using 0.3 mm inner-diameter PTFE tubes immersed in the water tank. Non-relevant background elements (top and bottom right) are blurred for clarity. *(c)* Diagram of the ultrasonic matrix array with 1 024 elements. The elements are divided into four separate panels along the y-axis. The origin of the coordinate system is indicated by the cross at the centre of the probe.

### 2.3. Acquisition sequences

The bimodal data acquisition was performed sequentially in two steps. In the first step, a 3D imaging sequence, referred to as the (PA+US) sequence, was performed to acquire PA volumes at multiple optical wavelengths together with their corresponding B-mode US volumes. In a second step, microbubbles were injected during a ULM acquisition. For *in vitro* experiments, microbubbles were delivered through continuous infusion, whereas bolus injections were used for *in vivo* experiments. The US probe was kept in the same position throughout both acquisition steps, ensuring that identical volumes of interest were imaged. The stability of the probe position was confirmed by comparing the US images acquired during the two steps.

#### 2.3.1. 3D photoacoustic imaging sequence

The (PA+US) sequence was built by combining acquisitions of PA data at a single optical wavelength and pulse-echo US data for each 3D volume. Each laser pulse provided optical excitation and simultaneously triggered a reception-only event on the US scanner, acquiring PA signals on one panel of the matrix probe (256 elements in parallel). To cover the four panels and record signals on all 1 024 probe elements, successive laser excitations were used. Pulse-echo US acquisitions were interleaved between laser pulses. The 50 ms interval between two laser pulses was divided into twelve 4.17 ms slots. PA acquisition occurred during the first slot, followed by a 4.17 ms pause. Nine slots were then used for US plane wave (PW) transmit-receive events (as described below), followed by another pause. The pauses immediately before and after the laser pulse prevented interference between PA-generated acoustic waves and those generated by PW emissions. B-mode images were formed by coherently compounding transmissions of nine US PWs steered at 0°, ±2.5°, and ±5° in both elevation and lateral directions (Figure 1(c)).

Each US pulse consisted of a burst of two cycles at a centre frequency of 7.8 MHz. To reconstruct a complete B-mode volume with the US scanner comprising 256 channels, the “light” inter-panel scheme was used, in which each panel receives backscattered signals resulting from transmissions performed by itself and its adjacent panels [26]. For example, panel 2 is active in receive mode when PWs are transmitted by panels 1, 2, and 3. As a result, each steered angle was transmitted ten times to reconstruct a complete volume. In summary, acquiring a 3D PA volume and a 3D B-mode volume at a single optical wavelength required 10 laser pulses and 90 PW US transmissions. For each panel, two PA signal recordings were processed. In summary, acquiring a 3D PA volume and a 3D B-mode volume at a single optical wavelength required 10 laser pulses and 90 PW US transmissions. For each panel, two PA signal recordings were processed. Consequently, eight PA recordings were used for image reconstruction, whereas the two redundant recordings were discarded.

For US acquisitions, signal recording began 4 µs after US transmission to allow the switches to transition from the transmit to the receive panel. In contrast, for PA acquisitions, the multiplexing switches selected the appropriate panel and were fully closed 30 µs before the laser pulse. Signal acquisition started simultaneously with the laser emission. The 30 µs time lag ensured electromagnetic compatibility between the recording electronics of the US scanner and the laser power supply during the laser-pulse discharge and enabled reliable recording of the PA signals.

PA and US signals were digitized at sampling rates of 62.5 MS $s^{-1}$ and 31.25 MS $s^{-1}$, respectively. The receive gains were kept constant across depth for both modalities. They were chosen to ensure sufficient signal amplitude for accurate digitization while avoiding saturation. Compared to the ULM acquisition, the (PA+US) sequence used a higher number of PW steered angles and retained more samples (lower decimation rate) for US imaging. Given the low PA acquisition rate and small data volume, image quality was prioritized over acquisition speed. The US data acquired during the (PA+US) sequence were beamformed to generate a B-mode image, providing tube positions for *in vitro* experiments and anatomical context for *in vivo* acquisitions.

For multispectral acquisitions, the laser wavelength was updated after every 10 laser pulses while the acquisition sequence continued uninterrupted. *In vitro*, optical excitation was performed at 16 wavelengths ranging from 700 to 850 nm in 10 nm increments. *In vivo*, five wavelengths (Section 2.1) were used for hindlimb imaging, whereas the full set of 16 wavelengths was retained for tail imaging. Consequently, a multispectral acquisition consisted of 50 optical excitations over 2.5 s when using five wavelengths, and 160 optical excitations over 8 s when using 16 wavelengths. To improve the signal-to-noise ratio (SNR), the complete multispectral (PA+US) acquisition sequence was repeated ten times for *in vitro* imaging and four times for *in vivo* imaging.

#### 2.3.2. 3D ultrasound localization microscopy sequence

The ULM acquisition sequence employed an ultrafast US acquisition sequence based on the coherent compounding of five US plane waves, steered at 0° and ±5° in both elevation and lateral directions (Figure 1(c)). Each pulse comprised two cycles and corresponded to an excitation centred at 7.8 MHz, with a 69% duty cycle and an approximate pulse repetition frequency of 14 kHz. The initial peak voltage (maximum voltage applied to the transducer) was set to maximize the received signals while preserving bubble integrity (6 V *in vitro* and 8 V *in vivo*). Received signals were sampled using the Vantage system's 100 % bandwidth sampling mode.

As in the (PA+US) sequence, the "light" inter-panel acquisition mode [42] was used, which requires ten emissions per steering angle to account for all transmit–receive combinations. With five steering angles, a total of 50 emissions were required to reconstruct each volumetric frame.

*In vitro*, a total of 100 batches of 200 volumetric frames were acquired at a volume rate of 285 Hz, corresponding to a total acquisition time of ~1.2 minutes and ~50 GB of raw radio-frequency data. *In vivo*, 500 similar batches were acquired at the same rate, yielding a total acquisition time of ~6.0 minutes and ~250 GB of radio-frequency data. Batch saving was performed in parallel with the acquisition of the subsequent batch and required up to 0.3 s per batch.

### 2.4. Micro-vessel imaging phantoms

#### 2.4.1 Vessel-mimicking phantom configurations

Blood vessel-mimicking phantoms suitable for both PA and ULM imaging were built using polytetrafluoroethylene (PTFE) tubes (internal diameter: 0.3 mm, wall thickness: 0.15 mm; S1810-04, Bola, Germany). This tube configuration had previously demonstrated robust photoacoustic performance in a calibrated PA spectrophotometer setup [39,43]. Its optical and acoustic transparency is therefore well suited to the operational range used here.

The 0.3 mm inner diameter corresponds to small arteries or large arterioles, providing a physiologically relevant microvascular model while ensuring compatibility with both imaging modalities. This tube diameter balances PA and ULM constraints. Although a smaller diameter would have improved PA reconstruction of the tube interior (filled with an absorbing solution), given the centre frequency of the US probe [44], it would have increased shear stresses during the microbubble solution injection for ULM and raised the risk of microbubble destruction. With the 0.3 mm diameter, microbubble integrity was preserved, though under these conditions, PAI predominantly revealed the tube boundaries rather than its full cross-section.

For acoustic coupling, the tubes and the US window of the probe were immersed in a water tank maintained at room temperature. Two uncoated microscope slides inserted on opposite sides of the tank served as optical windows. The tubes were secured within the tank using pairs of perforated plates positioned outside the illuminated volume. This configuration enabled precise and adaptable tube positioning, allowing the construction of phantoms with various configurations.

Four phantoms were designed. Phantom #1 used a 20 µm diameter black nylon thread (NYL02DS, VetSuture, Paris, France) placed at the volume-of-interest centre and oriented along the y-axis. This phantom enabled assessment of the spatial resolution of both PA and B-mode imaging, as well as estimation of the fixed acquisition delays required for beamforming. Phantom #2, a similarly centred PTFE tube, enabled evaluation of system sensitivity to blood-mimicking fluids, modality co-registration, and spectral unmixing. Phantom #3 featured two intertwined PTFE tubes aligned along the y-axis to demonstrate the complementarity of the two modalities. For Phantoms #1, #2, and #3, the thread or tubes were positioned quasi-parallel to the probe surface to mitigate the effects of the limited view in PAI, which results from the finite aperture of the matrix probe. Phantom #4 consisted of a PTFE tube coiled into a loop to compare the vessel detection sensitivity of PAI and ULM. The loop was secured with a white thread, which was not visible in PA images because of its low optical absorption and did not appear in ULM images because it contained no flowing microbubbles.

#### 2.4.2. Sulphate mixtures as photoacoustic agents

To evaluate the performance of our PA setup for mapping blood vessels and determining oxygen saturation, Phantoms #2, #3, and #4 were filled with solutions mimicking blood at distinct $SO_2$ values. These solutions were designed to (1) replicate the PA coefficient of whole blood in the NIR and (2) achieve concentration ratios reflecting physiological $SO_2$ levels [39]. For these solutions and for whole blood, the PA coefficient, $\theta^{PA}(\lambda)$, corresponds to the product of the optical absorption coefficient and the Grüneisen parameter normalized to water.

Specifically, the solutions consisted of aqueous mixtures of nickel(II) sulphate hexahydrate ($NiSO_4 \cdot 6H_2O$, an analogue for $HbO_2$) and copper(II) sulphate pentahydrate ($CuSO_4 \cdot 5H_2O$, an analogue for Hb). We denote $SO_2^{analog}$ as the ratio of sulphate species concentrations:

$$SO_2^{analog} = \frac{c^{Ni}}{c^{Ni} + c^{Cu}} \quad (3)$$

where $c^{\mathrm{Ni}}$ and $c^{Cu}$ correspond to the molar concentrations of $NiSO_4 \cdot 6H_2O$ and $CuSO_4 \cdot 5H_2O$, respectively.

$NiSO_4 \cdot 6H_2O$ (CAS: 10101-97-0, $M_w$ = 262.85 g mol$^{-1}$, ACS reagent, ≥98 %) and $CuSO_4 \cdot 5H_2O$ (CAS: 7758-99-8, $M_w$ = 249.69 g mol$^{-1}$, ACS reagent, ≥98 %) were purchased from Sigma-Aldrich (St. Louis, MO, USA). All solutions were prepared using analytical-grade water (ρ = 18 MΩ cm$^{-1}$, Purelab Option Q, ELGA LabWater). Stock solutions of $NiSO_4$ and $CuSO_4$ were prepared from the corresponding salts to obtain concentrations of 1.1 mol L$^{-1}$ and 0.36 mol L$^{-1}$, respectively. Mixtures were then prepared to mimic $SO_2$ levels of 60, 70, 80, and 95 % [39].

The optical absorbance spectra of the prepared solutions were measured using a UV-Vis-NIR spectrophotometer (VWR® P4 UV-Vis Spectrophotometer, VWR, Leuven, Belgium) in absorbance mode, using a 2 mm length quartz cuvette (QS 10.00 Hellma). The blank measurement was performed with analytical-grade water.

The prepared sulphate mixtures were injected into the tubes for subsequent acquisitions using the (PA+US) imaging sequence. For each acquisition, an initial acquisition with the tube filled with water was performed and used in post-processing to remove signal components unrelated to the injected solution (e.g., potentially stained tube walls). Additionally, each liquid was flushed out with air before the next one was injected.

### 2.5. Microbubble contrast agents for ultrasound localization microscopy

For the ULM acquisitions performed on the vessel-mimicking phantoms, blood perfusion with US contrast agents was mimicked by injecting the tubes with a solution of sulphur hexafluoride microbubbles (SonoVue®, Bracco, Milan, Italy). The microbubbles were prepared according to the protocol of the manufacturer and diluted in water at a ratio of 1:3,300. For each acquisition, lasting 1.2 minutes, 1 mL of this dilution was slowly and continuously injected into the PTFE tubes through needles with an outer diameter of 18 gauges to minimize shear stress.

### 2.6. Animal experiments

Mouse experiments were approved by the Animal Ethics Committee of Université Paris Cité and authorized by the French Ministry of Higher Education and Research and Space under reference number APAFIS #57772. All procedures were conducted in accordance with applicable French and European regulations on the protection of animals used for scientific purposes. Experiments were performed on BALB/c female mice aged 8 weeks (Janvier Labs, Le Genest-Saint-Isle, France). Animals were housed in a conventional animal facility under a 12 h light/12 h dark cycle at a controlled temperature of 22 °C, with food and water available *ad libitum*.

Mice were anesthetized throughout the experiment with 2 % isoflurane and maintained at 37 °C using a temperature-controlled heating pad. Before imaging, the hindlimb or the dorsal region at the base of the tail was depilated using a depilatory cream. A 27G catheter was inserted into the tail vein for intravenous administration of SonoVue® microbubbles during ULM acquisitions. The imaged volume was placed within the previously defined volume of interest (Section 2.2) for bilateral illumination. Transparent acoustic gel was applied between the tissue and the probe for acoustic coupling.

The 3D (PA+US) imaging sequence was first applied without contrast agent injection to perform multispectral PAI at five (hindlimb) or 16 (tail) optical wavelengths. Volumetric ULM acquisition was subsequently performed for approximately 6 min, during which SonoVue® microbubbles were injected intravenously as 25 µL boluses every minute. Animals were monitored until full recovery from anaesthesia after the experiment.

### 2.7. Post-processing

#### 2.7.1. Volumetric image reconstruction

Volumetric image reconstruction was performed using delay-and-sum (DAS) beamforming algorithms. The US signals were not pre-processed; instead, conversion from real to complex values was performed directly during reconstruction. The PA signals were normalized by the corresponding pyrometer measurements to compensate for pulse-to-pulse laser energy fluctuations and bandpass-filtered between 3 and 12 MHz using a third-order Butterworth filter. The filtered signals were then converted into the complex domain using the Hilbert transform.

One-way (PA) and two-way (US) propagation times between the probe elements and each voxel were calculated assuming a homogeneous medium with a constant speed of sound. For *in vitro* acquisitions, the speed of sound was set to that of water at the measured ambient temperature, whereas for *in vivo* acquisitions, it was fixed at 1 540 m $s^{-1}$. Voxel amplitudes were reconstructed by coherently summing the contributions from all array elements corresponding to the expected time-of-flight.

For both PA and US reconstructions, dynamic aperture apodization along the array surface was applied to maintain a constant angular aperture, using a f-number of 1.5 with a Welch window. The DAS algorithms consisted of linear operations applied to the signals and resulted in volumetric images with complex values. A reconstructed PA volume corresponded to eight laser pulses. A reconstructed US volume corresponded to 90 PW emissions in the (PA+US) mode and 50 PW emissions in the ULM mode.

Positive images were generated by computing the magnitude of the complex value at each voxel, resulting in envelope-detected images. Envelope detection was performed only as a final processing step, after all linear operations were applied to the complex-valued images. These operations included PA image averaging and, for *in vitro* experiments, background subtraction using a complex PA image acquired with water-filled tubes. The background and corresponding PA acquisitions were performed with the same imaging parameters and within a few minutes apart to minimize potential drift.

PA and conventional B-mode images were reconstructed on isotropic grids with voxel sizes corresponding to one-quarter of the acoustic wavelength. For ULM, the US data were first reconstructed on coarser isotropic grids with voxel sizes of half a wavelength, then refined to achieve a final 3D ULM resolution of one-twentieth of the wavelength. This subwavelength resolution was achieved through ULM-specific post-processing [45] (Section 2.7.4).

#### 2.7.2. Estimation of oxygen saturation from *in vitro* photoacoustic data

We estimated $SO_2^{analog}$ on a voxel-by-voxel basis from photoacoustic images.

**Per-wavelength calibration with a reference solution**

To compensate for wavelength-dependent variations in the average optical fluence delivered by the illumination system, voxel amplitudes were globally multiplied by a wavelength-dependent calibration factor. This factor was determined using a reference solution filling the tubes of phantom #2, #3, or #4: an aqueous nigrosine solution (CAS 8005-03-6, $M_w$ = 616.49 g $mol^{-1}$, high purity biological stain; Thermo Fisher Scientific, Waltham, MA, USA). In the NIR range, nigrosine converts absorbed optical energy into PA signals with the same Grüneisen coefficient as water [43]. The solution was prepared at 0.17 g $L^{-1}$, corresponding to an absorption coefficient (base *e*) of 2.0 $cm^{-1}$ at 760 nm.

At each wavelength, 10 PA acquisitions were averaged, and an acquisition with the tube filled with water was subtracted before envelope detection. The calibration factor was calculated as the ratio of the spectrophotometrically measured absorption coefficient of the reference solution to its mean PA amplitude within a predefined ROI. This ROI was identical for all wavelengths and was defined from the 700 nm PA image by selecting voxels with amplitudes exceeding 1.5 times the binarization threshold obtained using Otsu's method [46]. The resulting wavelength-dependent calibration factor was then applied to the PA images of the sulphate mixtures subsequently injected into the tubes.

**Estimation of $SO_2^{analog}$ from sulphate mixtures**

For each sulphate mixture, PA images were pre-processed by averaging 10 PA acquisitions, followed by background correction and envelope detection. Since $SO_2^{analog}$ is computed as a ratio (Equation 3), background voxels may yield arbitrary values unrelated to meaningful PA signals. Therefore, the tube was first delineated in 3D. The ROI was selected using the pre-processed PA image at 700 nm. A 3D Gaussian filter (σ equal to half the voxel size) was applied to smooth the image. The filtered image was binarized using a threshold equal to 1.5 times the threshold determined by Otsu's method, and connected components were extracted. A size threshold was then determined by applying Otsu's method to the distribution of connected-component sizes. Only voxels belonging to connected components larger than this threshold were retained for further analysis.

The wavelength-dependent calibration factors derived from the reference nigrosine solution were applied to the PA amplitudes of the pre-processed voxels within the ROI. $SO_2^{analog}$ was computed for each voxel of the ROI by first estimating the relative weights of $CuSO_4{\cdot}5H_2O$ and $NiSO_4{\cdot}6H_2O$ in the voxel. Weights were estimated by solving a linear system using a non-negative least-squares algorithm, with the molar absorption spectra of $CuSO_4{\cdot}5H_2O$ and $NiSO_4{\cdot}6H_2O$ as inputs. $SO_2^{analog}$ was computed as the ratio of weights, as defined in Equation 3, assuming proportionality between the estimated weights and compound concentrations. The resulting $SO_2^{analog}$ maps were overlaid onto the 700 nm PA images and visualized in 3D using Napari [47].

**2.7.3. Estimation of oxygen saturation from *in vivo* photoacoustic data**

The estimation of $SO_2$ from the *in vivo* PA data followed the same processing pipeline as that used for the *in vitro* experiments, except that no background subtraction was applied. Calibration factors were obtained with Phantom #2. The same spectral unmixing algorithm was applied to estimate the relative contributions of $HbO_2$ and Hb, from which $SO_2$ was computed (Equation 1).

**2.7.4. Subwavelength mapping of micro-vascularization in 3D**

The ULM data were analysed using an open-access pipeline [45]. The main processing steps are briefly described below, with the input parameters used in this study indicated in italics.

**Filtering.** Complex volumetric frames were filtered using singular value decomposition (SVD) to remove the tissue component and enhance the microbubble signal. The number of singular values removed was optimized separately for *in vitro* and *in vivo* datasets (15 of 200 singular values removed for the *in vitro* experiments and 20 of 200 for the *in vivo* experiments; "*min_svd_value*").

**Localization of local maxima.** The expected number of microbubbles per volume ("*number_of_particles*") was set to 130 for *in vitro* acquisitions with a single tube, 200 for acquisitions with two tubes, and 900 for *in vivo* acquisitions. The full width at half maximum ("*fwhm*") of the bubble peaks was taken equal to 5 × 5 × 5 voxels. Regional maxima were further filtered to ensure correspondence to microbubbles: only maxima with a local SNR greater than 9 dB ("*min_snr*") on sub-volumes of 7 × 7 × 7 voxels ("*patch_size*") were kept.

**Subwavelength localization.** For each localized maximum, a 3D radial symmetry approach was used to determine the coordinates of the microbubble centre with sub-wavelength precision.

**Tracking microbubbles.** The Hungarian algorithm was used to associate microbubble localizations between consecutive frames by minimizing the total Euclidean distance between localization pairs, thereby reconstructing the microbubble trajectories. The maximum inter-frame speed was set to 200 mm $s^{-1}$ (*in vitro*) and 100 mm $s^{-1}$ (*in vivo*) (“*max_velocity*”). Microbubbles were not allowed to disappear and reappear during tracking (*max_gap_closing* = 0). Tracks shorter than 10 (*in vitro*) or 15 (*in vivo*) detections (“*min_length*”) were discarded. The selected tracks were linearly interpolated by inserting one point between consecutive localizations. Tracks were then discretized on a grid 10 times finer than that of the input US data (“*res*”).

**Micro-vascularization maps.** From the interpolated tracks, three 3D renderings were generated: density, with voxel values equal to the number of tracks passing through each voxel; velocity, corresponding to the magnitude of the velocity vector; and directivity, obtained by assigning a positive or negative sign to the speed according to the predominant direction of propagation along the depth axis.

The 3D renderings were displayed using maximum amplitude projection (MAP) along the coordinate system axes.

## 3. Results

### 3.1. Spatial resolution and co-registration of the photoacoustic and ultrasound images

Figure 2 displays images acquired using the 3D (PA+US) imaging sequence for Phantom #1 and Phantom #2. Both phantoms contain an elongated structure oriented along the y-axis, with a similar position relative to the imaging system: a 20 µm diameter thread (Phantom #1) and a 0.3 mm inner diameter tube (Phantom #2). This similarity in position, combined with the difference in structure size, allows characterization of the PAI performance for vessel-mimicking structures.

Figure 2(a) shows PA and US cross-sectional images at y = 1.4 mm for Phantom #1, corresponding to the plane with the highest mean amplitude. The displayed PA image was acquired at 700 nm. The black nylon thread (~20 µm) is about one order of magnitude smaller than the acoustic wavelength at the centre frequency of the array (190 µm), modelling a point-like optical absorber and acoustic scatterer in the xz-sections. Phantom #1 was used to estimate fixed hardware-induced temporal offsets. These offsets were incorporated into the reconstruction algorithm for all other acquisitions, ensuring sharp image reconstruction and axial co-registration of PA and US images. In both modalities, the thread appeared as a single 2D Gaussian-like spot in cross-sectional images. A 2D Gaussian function was fitted independently to the PA and US spots to estimate their centroid coordinates and full width at half maximum (FWHM), enabling assessment of spatial resolution and co-registration accuracy:

$$A(x) = A_0 \cdot \exp\left(-\frac{(x-x_0)^2}{2\sigma_x^2} - \frac{(z-z_0)^2}{2\sigma_z^2}\right) \tag{4}$$

where $A_0$ is the amplitude, $x_0$ and $z_0$ the centroid coordinates, and $\sigma_x$ and $\sigma_z$ the standard deviations along x and z. The FWHM was computed as $FWHM_i = 2\sqrt{2\ln(2)} \times \sigma_i$ to estimate lateral and axial resolutions. The inter-centroid distance between PA and US spots was calculated as $d_{PA\text{-}US} = \sqrt{({x_0}^{PA} - {x_0}^{US})^2 + ({z_0}^{PA} - {z_0}^{US})^2}$. For robustness, the estimated parameters were averaged over 10 independent image reconstructions, 14 cross-sections (y = −1.5 to 5.0 mm in 0.5 mm increments), and 16 optical wavelengths. Mean values and standard deviations are reported in Table 1.

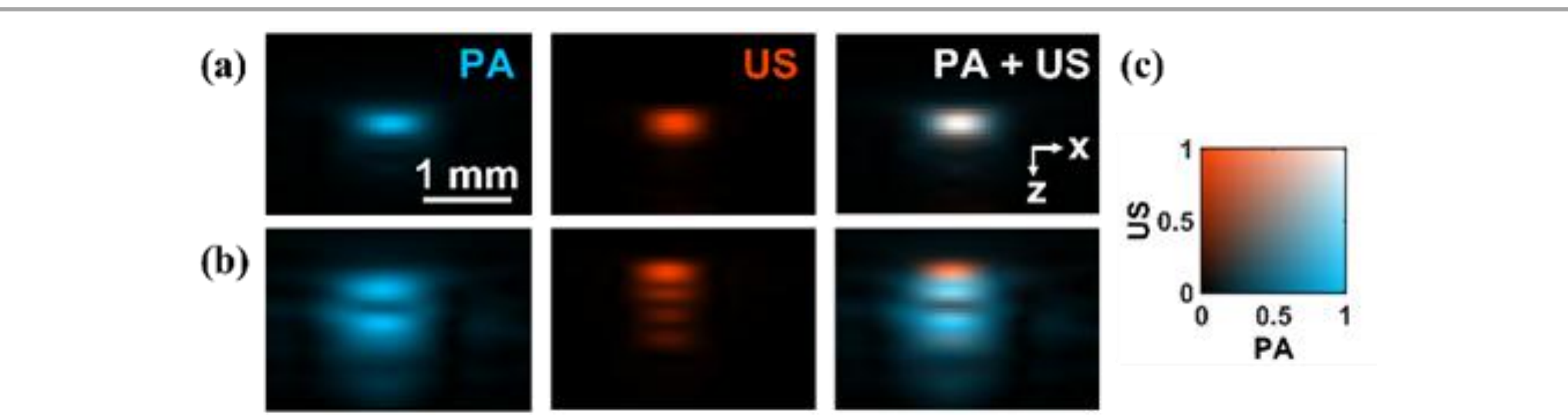

**Figure 2.** Cross-sectional images at y = 1.4 mm from Phantom #1 *(a)* and Phantom #2 *(b)*. In Phantom #2, the 0.3 mm tube is filled with a sulphate mixture ($SO_2^{analog}$ = 60 %). Each image covers a field of 3.0 mm (x-axis) × 2.0 mm (z-axis) and is centred at (x = 0.0 mm, z = 5.6 mm). The photoacoustic (PA, left; optical excitation at 700 nm) and ultrasound (US, centre) images are displayed on a linear scale. The right panels show the overlays. Panel *(c)* shows the colour map.

The inter-centroid distance along z was expectedly small due to temporal offset adjustment. The global inter-centroid distance remained small relative to the system spatial scales: approximately 8 % of the US wavelength, half the voxel size (25 µm *vs.* 50 µm), and less than 6 % of the FWHM. Low standard deviations (relative to system scales) indicate minimal variability across wavelengths, iterations, and y positions, confirming excellent co-registration. Axial resolution is on the order of the US wavelength for both modalities, while lateral resolution is consistently lower. Axial resolution is determined by the signal bandwidth, whereas lateral resolution is diffraction-limited and affected by the limited numerical aperture of the probe and the directivity of its elements. The lower lateral resolution in PAI, visible as a blue halo in the overlaid image (Figure 2(a), right panel), arises from two main factors: broader and richer low-frequency content in PA signals and the absence of emission focusing (PA focuses only on receive; US benefits from both emission and receive focusing).

Figure 2(b) shows cross-sectional images of Phantom #2 (0.3 mm tube filled with a sulphate mixture). Unlike Phantom #1, the tube appears as multiple spots along the axial direction due to its finite size, dimensions larger than the US wavelength, and geometry. US imaging highlights the tube walls through specular echoes, as both tube walls and the interior are homogenous and do not scatter. The upper spot corresponds to direct echoes on the exterior wall, similar to Phantom #1; lower spots may result from direct echoes or multiple reflections. As the speed of sound inside the walls was not used in the reconstruction process, the spot positions could not be used to measure distances. Side walls were not visible due to the limited aperture of the array and PW incident angles.

For PAI, a background image acquired with the tube filled with water was subtracted before envelope detection. The optically absorbing portion of the phantom appeared as two distinct axial spots. With an absorption coefficient of approximately 4 cm$^{-1}$, the tube was optically thin, resulting in nearly homogeneous illumination of the solution across the xz-plane. Given the tube inner diameter (0.3 mm), the generated PA signal was expected to exhibit a first and dominant spectral peak around 2 MHz, with an upper bound near 6 MHz [44]. If fully detected, this spectrum would enable reconstruction of the entire tube cross-section using the DAS algorithm. However, because the transducer bandwidth captured only part of the generated spectrum, the reconstructed image mainly highlighted the boundaries of the absorbing solution, leaving a gap in the tube interior.

**Table 1.** Experimental assessment of spatial resolution and co-registration accuracy.

| | Spatial resolution (FWHM[a], µm) | | Inter-centroid distance (µm) |
|---|---|---|---|
| | PA | US | |
| Along lateral axis (mean ± std) | 768 ± 57 | 466 ± 42 | 15 ± 25 |
| Along axial axis (mean ± std) | 309 ± 42 | 260 ± 32 | |

[a] Full width at half maximum.

The measured axial distance between the two PA spots (348 µm; Figure 3(c); n = 656 cross-sections) is consistent with the reconstruction of the tube's inner boundaries. The upper and lower tube walls are separated, as expected, given the system axial resolution (Table 1). The upper US spot, corresponding to the outer tube wall, was located above the upper PA spot, which corresponds to the inner wall. The tube side walls were not reconstructed because of limited-view effects. Despite these artifacts, the PA amplitude remained proportional to the absorption coefficient of the solution.

### 3.2. Bimodal imaging: co-registered photoacoustic and ultrasound localization microscopy images

After validation of the co-registration accuracy of the (PA+US) imaging sequence, the performance of the co-registered PA-ULM bimodal imaging was evaluated using Phantom #2. The tube was filled with a sulphate mixture corresponding to $SO_2^{analog}$ = 60 % for the (PA+US) acquisitions, followed by a ULM acquisition with microbubble injection.

Figure 3(a) shows the ULM density map superimposed on the volumetric PA image acquired at 700 nm, displayed as MAPs over a 0.3 mm thick slice along the lateral (x) and elevational (y) axes. A MAP over the entire volume is presented in Figure 4(a). The cylindrical geometry of the phantom is clearly recovered in the ULM reconstruction, and the two-spot PA pattern previously described in Section 3.1 is consistently observed over the full elevational extent. In PAI, the blood-mimicking solution is clearly detectable, indicating sufficient sensitivity for vascular imaging. Variations in PA amplitude along the tube were observed, including a lower signal between -5 mm and -2 mm along the y-axis, attributed to a slight tube tilt and non-uniform illumination; similar effects were also observed for Phantom #1. Periodic modulations of the PA signal along the elevational direction are likely caused by the kerf between transducer panels (Figure 1(c)). In contrast, the ULM reconstruction exhibits spatial continuity along the entire tube.

In each cross-sectional plane, the tube interior mapped by ULM was axially bounded by the two PA signal peaks corresponding to the inner tube boundaries (Section 3.1). To quantitatively assess the co-registration between the two modalities, which differ in spatial resolution and contrast mechanisms, the tube centroid in each cross-section was used as a modality-independent descriptor of tube position.

For ULM, the 3D density map was binarized to retain voxels containing at least one detected track. Morphological closing (using a spherical structuring element with a radius of four voxels) was applied to suppress discontinuities, followed by an opening (radius eight voxels) to remove isolated trajectories outside the tube. Internal cavities were then filled using 26 connectivity to obtain a contiguous binary mask of the tube. In each xz-cross-section, the centroid of the largest connected component was computed and defined as the ULM tube centroid. The equivalent diameter of this connected component provided an estimate of the tube diameter.

For PA images, the averaged volumetric image acquired at 700 nm was interpolated onto the ULM grid and binarized using a threshold equal to 2.5 times Otsu's threshold. In each cross-section, the centroid of the largest connected component was first computed. To account for the characteristic two-spot axial PA pattern, the axial amplitude profile passing through this centroid was extracted (Figure 3(c)), and the positions of the two peaks were identified. The PA tube centroid was then defined as the midpoint between these two peaks, while their axial separation provided an estimate of the tube diameter.

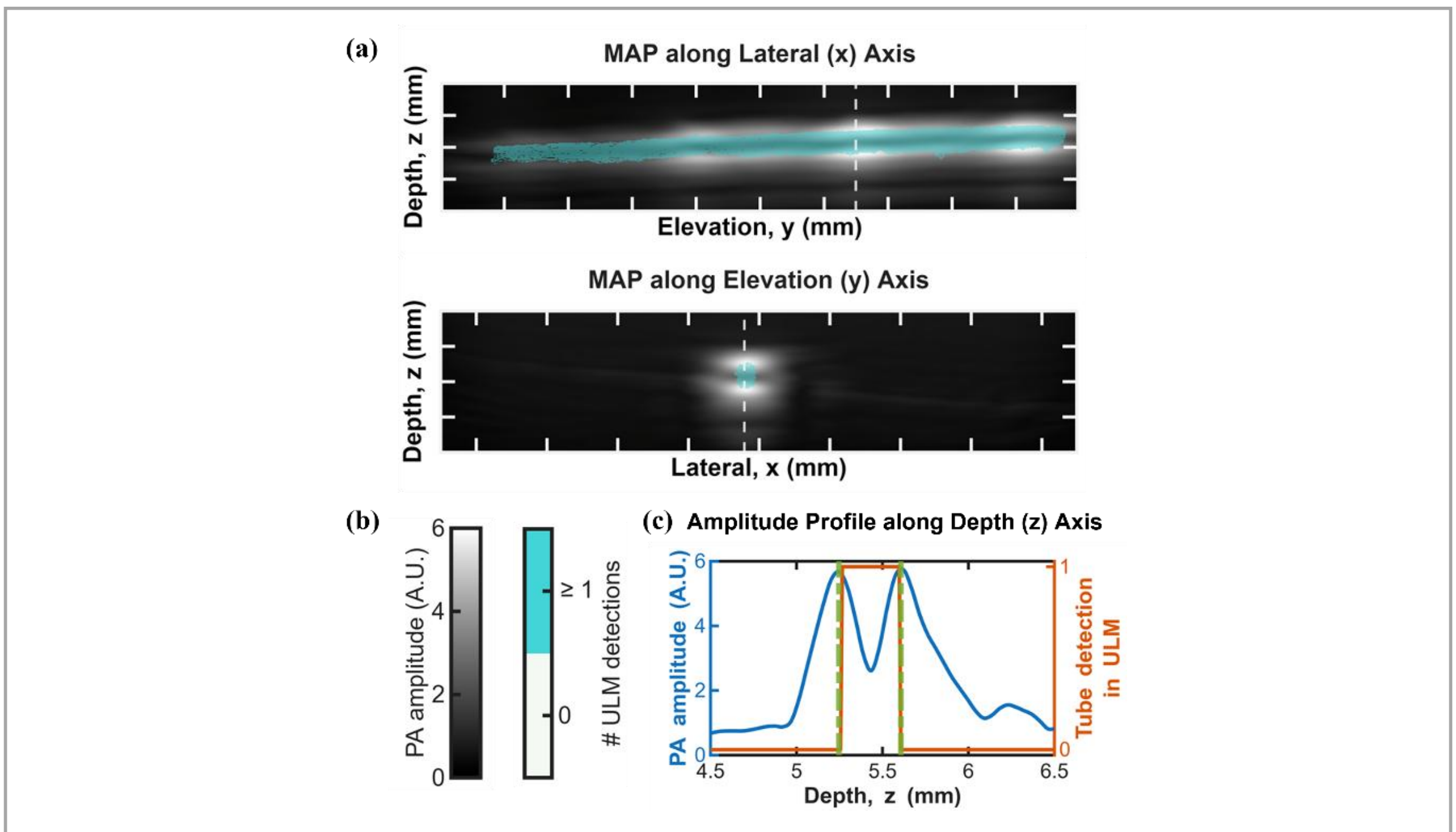

**Figure 3.** *(a)* Volumetric images obtained on a 0.3 mm tube filled with a sulphate mixture ($SO_2^{analog}$ = 60 %) by photoacoustic imaging at 700 nm (grayscale) and ULM (density map). The PA image is displayed on a linear scale; the density map is binary (cyan if at least one bubble detection in a given voxel). The two images correspond to maximum amplitude projections (MAP) along the lateral (top, $MAP_x$) and elevation (bottom, $MAP_y$) axes. $MAP_x$ is 9.9 × 2.0 mm$^2$ with the projection computed over x = -0.3 mm to x = 0.0 mm; $MAP_y$ is 9.0 × 2.0 mm$^2$ with the projection computed over y = 1.5 mm to y = 1.8 mm. Horizontal graduation: 1.0 mm. Vertical graduation: 0.5 mm. The colour bars are shown in *(b)*. *(c)* Detection of tube boundaries in a depth (z) profile at x = -0.2 mm and y = 1.5 mm (marked with white dotted lines in (a)) for both modalities. The green dotted lines indicate the peaks on the PA amplitude profile (solid blue line). The orange rectangular function shows the contours for ULM.

Using this approach, tube centroids were successfully determined for both modalities in 656 of 1,050 cross-sections. The Euclidean distance between PA and ULM centroids across these sections was 20 µm ± 11 µm (mean ± SD), indicating a high degree of spatial concordance between the two modalities and confirming the excellent co-registration accuracy of the proposed PA-ULM bimodal framework. Tube diameters were measured as 348 µm ± 14 µm in PA and 328 µm ± 30 µm in ULM, both consistent with the nominal internal diameter of 0.3 mm.

### 3.3. Complementarity of functional information

In PAI, the acoustic wavefields generated by elongated absorbers, such as tubular structures, are highly directional. Consequently, reconstruction requires at least one transducer element to be oriented approximately normal to the absorber axis for the structure to be visible in the reconstructed image [48]. In Phantom #2, the tube was straight and only slightly tilted (<3°) relative to the probe surface, a configuration favourable for PA detection. As a result, the tube was fully visible in the 3D PA reconstruction (Figure 4(a)). This limitation does not apply to ULM, which relies on isotropic scattering from microbubbles and is therefore not expected to suffer from limited-view artifacts.

To investigate the impact of tube orientation on visibility in the PA-ULM framework, Phantom #4 was introduced. This phantom consisted of a looped tube tilted by approximately 15° relative to the probe acoustic window. Figure 4(b) shows the corresponding PA and ULM volumetric reconstructions. In PAI, the tube was only partially visible, with high PA amplitudes mainly observed in the upper and lower portions of the loop (white solid arrows). These regions were more favourably oriented with respect to the detection geometry. In contrast, ULM revealed additional tube segments that were weakly visible or not detectable in PAI. However, the loop was still not fully reconstructed in ULM, as acoustic shadowing by the tube itself limited microbubble detection at greater depths (orange dotted arrow).

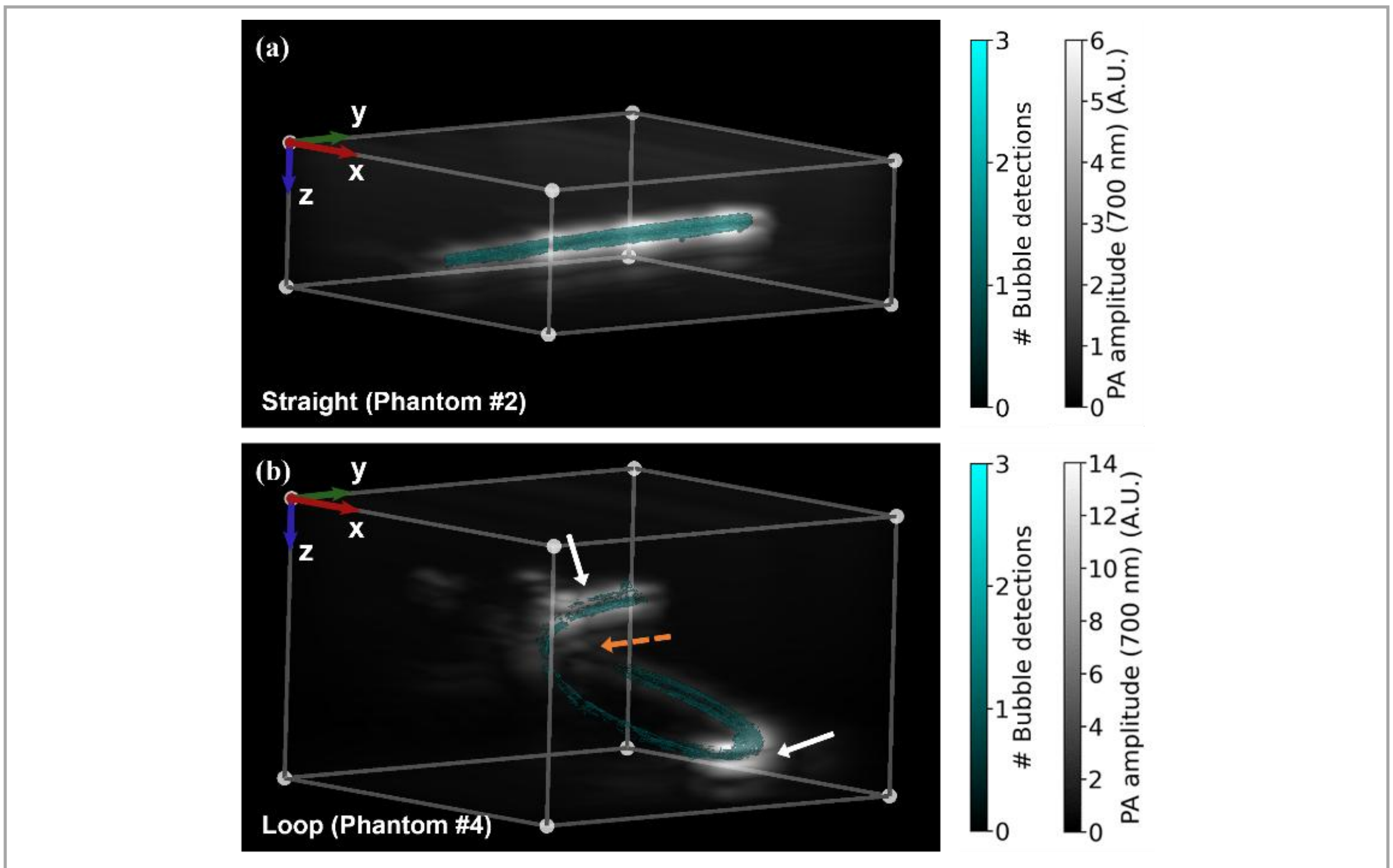

**Figure 4.** Bimodal volumetric images acquired on a 0.3 mm tube, either straight (*a*, Phantom #2) or looped (*b*, Phantom #4). The tube was filled with a sulphate mixture ($SO_2^{analog}$ = 60 % and $SO_2^{analog}$ = 95 % for Phantoms #2 and #4, respectively). The PA amplitude at 700 nm is shown in grayscale, while the ULM-derived density map is overlaid in cyan. The PA image and density map are displayed on linear scales. The top volume spans Δx × Δy × Δz = 9.0 × 9.9 × 3.2 mm$^3$ and is centred at (x, y, z) = (0.0, 0.0, 5.1 mm). The bottom volume spans 9.0 × 9.9 × 6.2 mm$^3$ and is centred at (0.0, 0.0, 6.4 mm).

Beyond their complementarity in terms of directivity, the two modalities also provide complementary contrast mechanisms and functional information. PAI maps optically absorbing structures and enables discrimination between absorbers, such as $HbO_2$ and Hb, based on their optical absorption spectra. ULM reconstructs vascular morphology and provides quantitative information on blood flow at the micrometre scale. This complementarity was further illustrated using Phantom #3, which consisted of two intertwined tubes. For PAI, both tubes were respectively filled with sulphate mixtures mimicking blood such as $SO_2^{analog}$ = 95 % and $SO_2^{analog}$ = 70 %, corresponding to arterial and venous conditions in humans. For ULM acquisitions, each tube was continuously and slowly injected with 1 mL of diluted microbubble suspension. Figure 5 shows the top-view velocity (a) and flow directivity (b) maps derived from ULM. Figure 5(c) presents the PA volume acquired at 700 nm, while Figure 5(d) presents the $SO_2^{analog}$ map obtained after spectral unmixing of the 16 wavelengths.

Both modalities clearly resolve the two tubes. In ULM, distinct flow dynamics are observed, with higher microbubble velocities in one tube compared to the other, and opposite predominant flow directions relative to the probe. In PAI, spectral unmixing successfully discriminates the two injected solutions, yielding $SO_2^{analog}$ values in close agreement with the expected concentrations. It should be noted that the tube-segmentation procedure used for $SO_2^{analog}$ estimation excluded regions with the weakest PA signals, namely between -5 mm and -2 mm along the elevational axis, resulting in sparser data in this region.

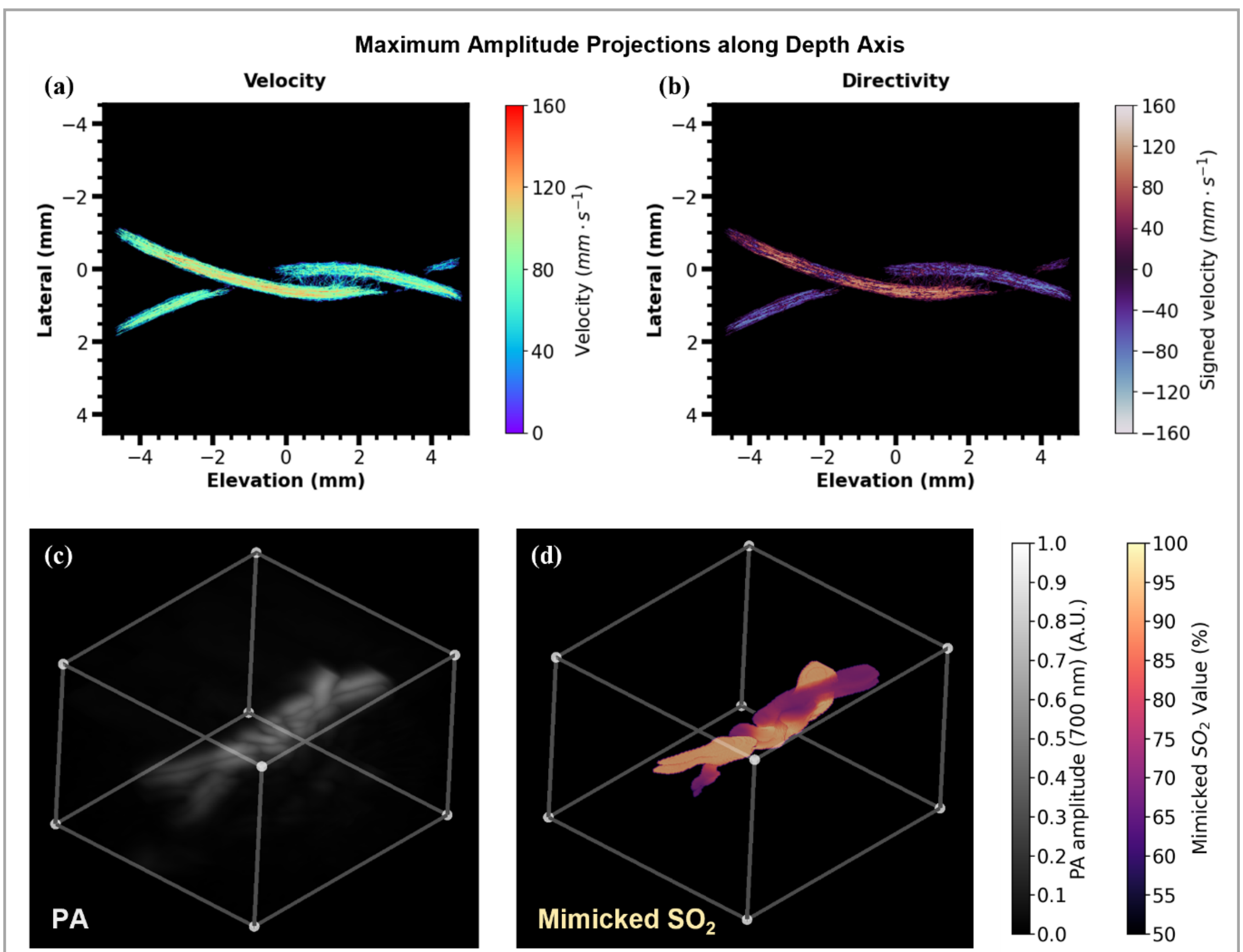

**Figure 5.** 3D maps of velocity *(a)* and flow directivity *(b)* obtained using ULM in two intertwined tubes with 0.3 mm inner diameters. Both images represent MAPs along the depth axis. *(c)* Volumetric PA image acquired at 700 nm (grayscale). *(d)* Corresponding $SO_2$ map (purple to yellow) obtained after spectral unmixing of 16 PA images acquired at optical wavelengths ranging from 700 to 850 nm. The tubes were filled with solution exhibiting $SO_2^{analog}$ values of 70 % and 95 %.

Finally, some tube segments were partially missing in both modalities. In ULM, this was mainly due to acoustic shadowing when the two tubes overlapped axially. In PAI, spectral colouring occurred because the optical fluence was attenuated after passing through the first tube before reaching the second.

### 3.4. Evaluation of the accuracy of spectral unmixing

As reported in the previous section, spectral unmixing applied to Phantom #3 yielded $SO_2^{analog}$ estimates in good agreement with the expected values. To quantitatively assess the accuracy of the spectral unmixing process, we analysed PA data acquired from Phantom #2, in which four sulfate mixtures with predefined oxygenation levels ($SO_2^{analog}$ = {60, 70, 80, 95} %) were injected.

Spectral unmixing was performed using two wavelength configurations: (i) the complete set of 16 wavelengths (700-850 nm, 10 nm increments), and (ii) the reduced set of five wavelengths (700, 750, 790, 820 and 850 nm; [39]). Table 2 reports the resulting $SO_2^{analog}$ estimates expressed as the mean value over all voxels within the tube, together with the corresponding standard deviations.

**Table 2.** Sulphate ratios obtained from spectral unmixing of photoacoustic data, considering either 16 or five wavelengths. The results are expressed as (mean value ± standard deviation), considering the sulphate ratios inside the tube.

| True $SO_2^{analog}$ (%) | Spectral unmixing of 16 wavelengths (%) | Spectral unmixing of 5 wavelengths (%) |
|---|---|---|
| 59.8 | 62.9 ± 1.7 | 64.8 ± 4.3 |
| 69.5 | 72.2 ± 4.6 | 72.9 ± 5.1 |
| 79.8 | 80.6 ± 0.9 | 81.1 ± 1.1 |
| 94.7 | 94.5 ± 0.4 | 94.6 ± 0.5 |

For the configuration with 16 wavelengths, the absolute deviation from the expected $SO_2^{analog}$ values ranged from 0.2 to 3.1 percentage points, with the maximum deviation observed at $SO_2^{analog}$ = 60 %. Using the configuration with five wavelengths, the deviation increased for all but the highest oxygenation level ($SO_2^{analog}$ = 95 %), reaching a maximum of 5.0 percentage points at $SO_2^{analog}$ = 60 %. For both wavelength configurations, the estimation error decreased with increasing $SO_2^{analog}$, and differences between the two configurations became negligible at $SO_2^{analog}$ = 95 %.

Overall, the use of 16 wavelengths resulted in consistently lower estimation errors compared to the configuration with five wavelengths, confirming the benefit of increased spectral sampling. Although reduced spectral sampling leads to increased estimation errors, the impact remains limited within the clinical physiological range, as both wavelength configurations exhibit their lowest error at $SO_2^{analog}$ = 95 %.

### 3.5. *In vivo* validation of the PA-ULM system for vascular imaging in healthy mice

To further validate our bimodal PA-ULM imaging system, *in vivo* imaging experiments were performed on healthy mice in regions with well-characterized vascular anatomies: the tail and the hindlimb.

Figure 6 shows representative results obtained from the tail of a healthy mouse. MAPs of (a) the PA image acquired at 800 nm (close to the isosbestic point of $HbO_2$ and Hb), (b) the ULM-derived density map, and (c) their overlay demonstrate a close spatial correspondence between the vascular structures detected by both modalities. The PA image reveals two blood vessels oriented parallel to the probe, consistent with the vessel orientation investigated in Phantom #2 and with the tail vascular anatomy. From PA images acquired at 16 wavelengths, $SO_2$ was estimated at three different (x, y, z) locations: (1) $SO_2^1$ = 76 % at (-0.6, -3.4, 7.2) mm; (2) $SO_2^2$ = 32 % at (0.6, 3.8, 5.8) mm; and (3) $SO_2^3$ = 92 % at (-1.0, 1.6, 6.9) mm. Using only the selected subset of five wavelengths (given in Section 2.1) yielded $SO_2$ estimates of 78 %, 35 %, and 93 % at locations (1), (2), and (3), respectively. As observed *in vitro* (Section 3.4), $SO_2$ values obtained using 16 and five wavelengths are consistent, with differences below 3%. These results confirm the robustness of *in vivo* $SO_2$ estimation and the suitability of the selected wavelengths. The measured $SO_2$ values are also consistent with the expected physiology of the mouse tail [49], suggesting that location (2) corresponds to a caudal vein, whereas locations (1) and (3) most likely correspond to the ventral caudal artery. The unfiltered ULM density map reveals the associated microvascular network dominated by the two main parallel vessels. Some discontinuities are observed along the ULM vessel trajectories, coinciding with regions of reduced PA signal intensity, suggesting that both modalities are similarly affected by local signal loss.

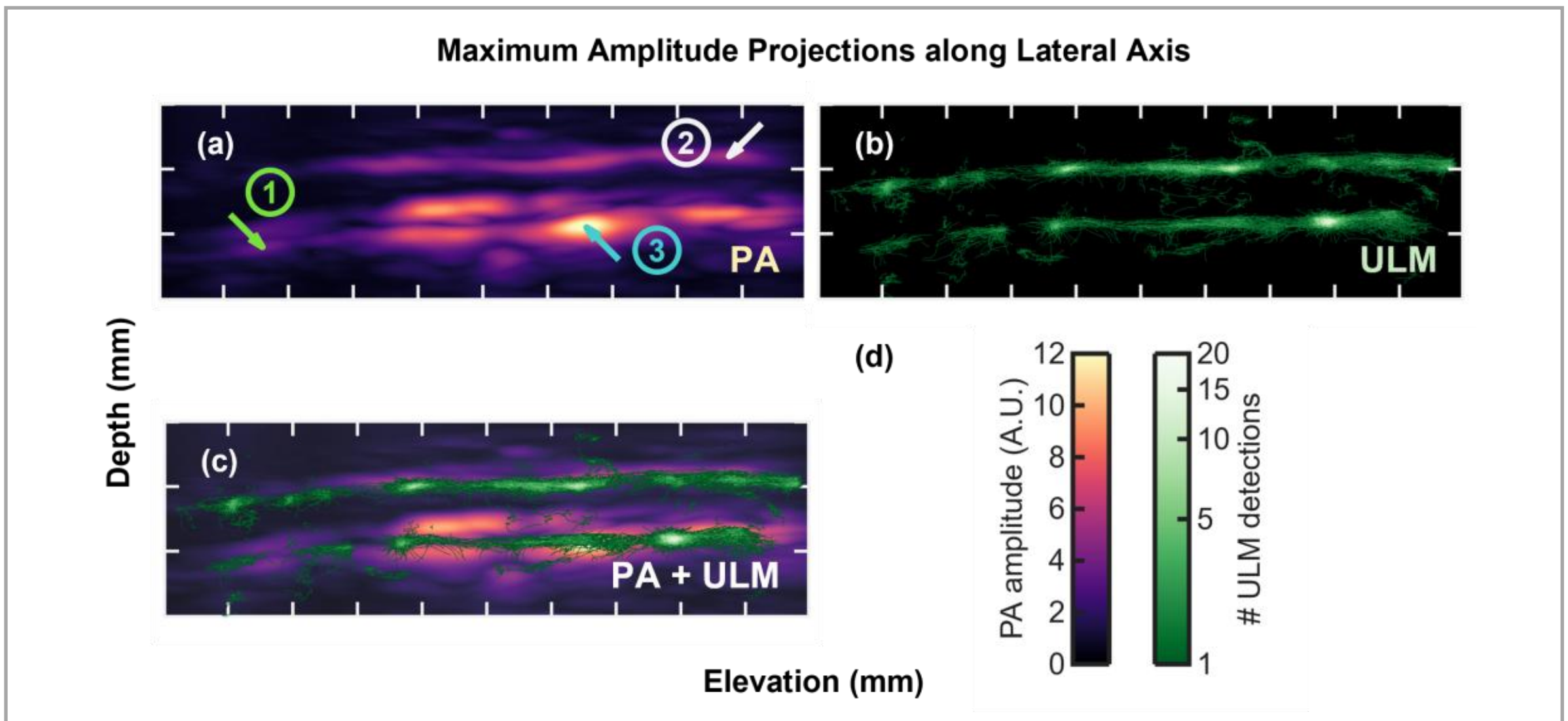

**Figure 6.** *(a)* Three-dimensional photoacoustic image of a mouse tail acquired at 800 nm and displayed as the maximum amplitude projection (MAP) along the lateral axis. The MAP spans a field of view of 9.9 × 3.0 $mm^2$ and was computed over the lateral range from x = -4.5 mm to x = 4.5 mm. Horizontal and vertical scale bars: 1.0 mm. The photoacoustic image is displayed on a linear scale. The three coloured arrows indicate the locations at which oxygen saturation values are reported in the main text. *(b)* Lateral MAP of the microvascular density map reconstructed using ULM. The image is displayed on a logarithmic scale. *(c)* Overlay of (a) and (b). *(d)* Colour bars for the photoacoustic and ULM images. The ULM values correspond to the number of detected tracks per pixel.

The vascular network of the mouse hindlimb exhibits a broader range of vessel orientations than the tail, providing a more challenging imaging configuration for PAI (Figure 7). Similar to Phantom #4, some vascular structures detected by ULM are not visible in the PA image because of the directional sensitivity of PAI. Nevertheless, the vascular structures identified by both modalities remain in close spatial agreement, supporting the robustness of the intrinsic co-registration under *in vivo* conditions. From PA images acquired at five wavelengths, $SO_2$ was estimated at three (x, y, z) locations: (1) $SO_2^1$ = 43 % at (-0.6, -1.6, 5.9) mm; (2) $SO_2^2$ = 72 % at (3.7, -1.4, 9.3) mm; and (3) $SO_2^3$ = 47 % at (-2.0, 4.2, 9.4) mm. These results are consistent with the $SO_2$ values reported for the tail, with location (2) corresponding to arterial blood and locations (1) and (3) to venous blood.

Overall, the *in vivo* experiments demonstrate that the proposed PA-ULM imaging system simultaneously provides detailed co-registered structural and functional information. ULM enables high-resolution visualization of the microvascular network beyond the spatial resolution and angular sensitivity of PAI, while multispectral PAI quantifies blood oxygen saturation, highlighting the complementary strengths of the two modalities for vascular imaging.

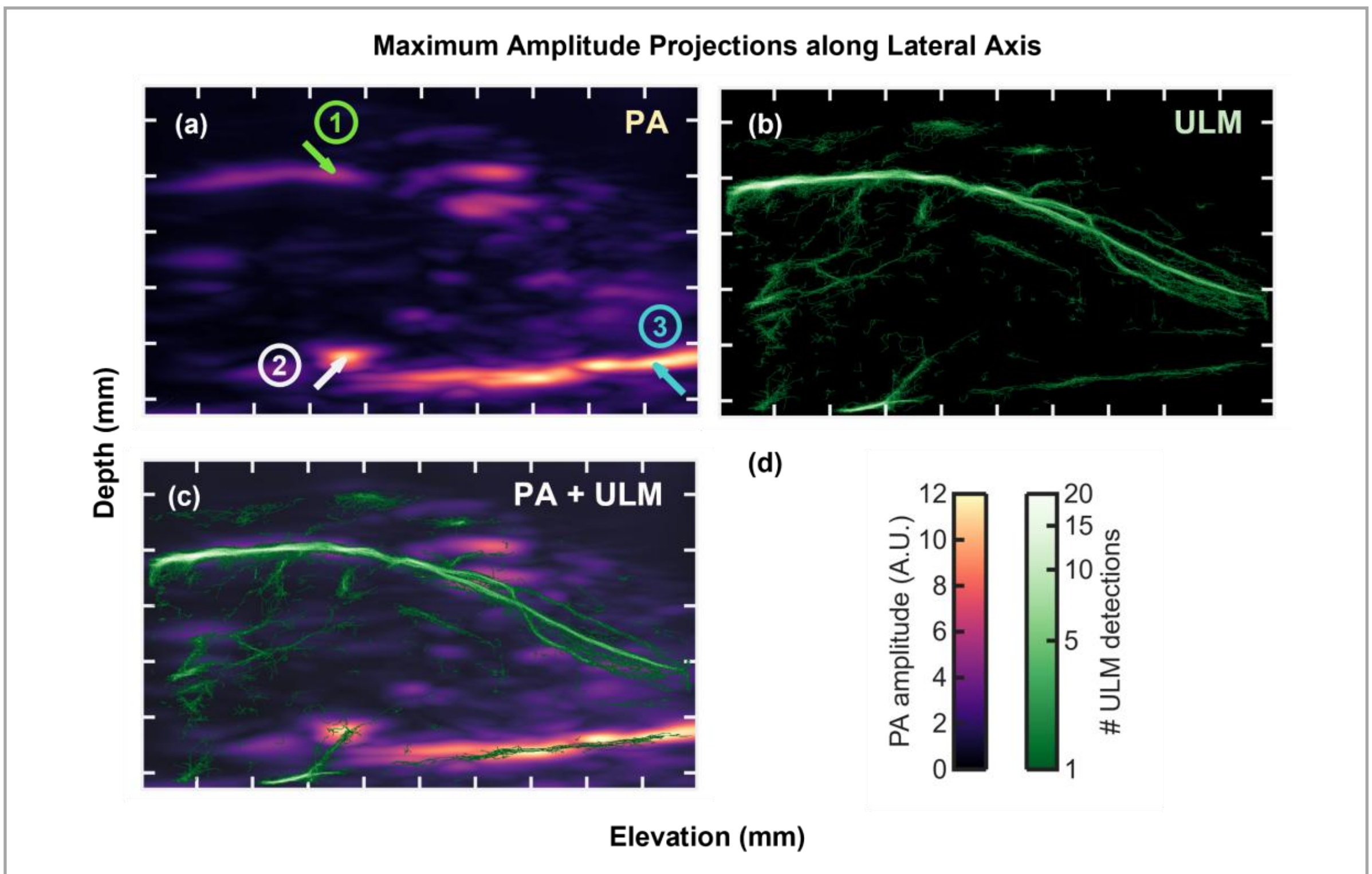

**Figure 7.** Three-dimensional images acquired from a healthy mouse hindlimb. The MAPs span a field of view of 9.9 × 5.9 mm² and were computed over the lateral range from x = -4.5 mm to x = 4.5 mm. All other display parameters are identical to those in Figure 6.

## 4. Discussion

We report the performance of a planar fully sampled US matrix array for combined 3D ULM and multispectral PAI, aiming to obtain intrinsically co-registered volumetric maps of tissue oxygenation and microvascular architecture. Our results demonstrate that a 7.8 MHz matrix array initially developed for volumetric ULM can also provide reliable multispectral PAI while preserving accurate spatial correspondence between both modalities. *In vitro* experiments demonstrated accurate volumetric co-registration between PAI and ULM and reliable vessel size estimation. Using *in vitro* vessel phantoms injected with blood-mimicking fluids, we further validated the capability of the system to perform multispectral PAI over up to 16 optical wavelengths and to estimate oxygen saturation with a maximum error of 5 %. *In vivo* experiments further validated the ability of the approach to provide co-registered maps of microvascular perfusion and oxygenation in healthy mice using anatomically well-characterized vascular structures. Together, these results establish the feasibility of a single-probe volumetric PAI/ULM approach for simultaneous structural and functional characterization of the microvasculature.

Despite these advantages, the current implementation also revealed limitations inherent to the use of planar US arrays for PAI. Pronounced limited-view artifacts were observed, resulting from both the restricted angular aperture of the array and the acoustic properties of the detectors. Compared with spherical arrays specifically designed for PAI [25,50], the relatively high centre frequency and limited bandwidth of the matrix array reduce sensitivity to a broad range of photoacoustic frequencies. Consequently, the current configuration is primarily suitable for vascular structures with favourable orientations relative to the array surface and within a limited range of vessel sizes.

Although ULM is also affected by limited-view effects, these mainly arise from incomplete microbubble localization and tracking when bubbles are not detected or when their trajectories are partially observed. In contrast, the limited-view artifacts encountered in PAI are directly related to the directional emission of acoustic waves generated by optical absorption in blood vessels. Several approaches could help mitigate these effects. Multispectral fluctuation PAI has recently demonstrated the ability to recover blood oxygenation and full-view vascular information by exploiting PA signal fluctuations induced by red blood cell motion [51]. However, this approach remains sensitive to tissue motion and noise, which requires further development. On the reconstruction side, the present study used a basic PAI reconstruction algorithm without incorporating prior anatomical information. Because ULM provides a high-resolution map of the vascular network, integrating ULM-derived structural information into PA reconstruction could potentially mitigate limited-view artifacts [52].

Another limitation of the current implementation is related to optical illumination geometry. Bilateral illumination enabled efficient deposition of optical energy throughout the volume of interest, but the mechanical support required for this configuration restricted accessibility in one dimension (lateral direction here). Alternative illumination strategies, such as top-side illumination [31] or matrix arrays incorporating a central opening [53], could enlarge the accessible field of view and open the possibility of handheld PAI/ULM probes. This would allow free orientation to minimize limited-view artifacts or to perform multi-view PAI acquisitions [54].

Beyond hybrid PAI/ULM imaging, localization optoacoustic tomography (LOT) has recently emerged as a photoacoustic counterpart to ULM, using highly absorbing nanoparticles to achieve super-resolution vascular imaging [55,56]. LOT may provide perfusion and oxygenation information using a single imaging modality. However, compared with ULM, LOT remains at an earlier stage of development and faces several important challenges. The first challenge will be the development of suitable PA contrast agents with sufficient performance and clinical translation potential, whereas ULM relies on clinically approved US contrast agents. A second challenge is penetration depth. Direct comparisons between LOT and ULM for mouse brain imaging through the intact skull have shown that LOT can provide higher SNR for superficial cortical imaging, particularly in older animals with thicker skulls, while ULM achieves greater penetration depth and enables whole-brain coverage [30]. Finally, the presence of a strongly absorbing exogenous contrast agent may interfere with endogenous haemoglobin signals and hinder quantitative oxygenation measurements.

Neither spherical US arrays developed for PAI nor planar US arrays used for ULM can achieve optimal performance in the complementary modality, due to the different acoustic requirements of the two techniques. Therefore, hybrid arrays combining a spherical array for PAI and a planar array for ULM could bring together the state-of-the-art performance of each modality and overcome their respective limitations. Hybrid arrays combining a linear array positioned at the centre of a spherical array have already been reported [57,58], demonstrating the feasibility of this approach and representing a first step toward volumetric PAI and ULM systems that address the physical constraints of both modalities. Such systems could ultimately enable comprehensive characterization of microvascular structure and function in biological processes where vascular remodelling and oxygenation changes are closely interconnected.

## 5. Conclusion

This work addresses the significant challenge of using a commercially available, single planar matrix array to provide intrinsic volumetric co-registration of ULM and PAI. Through rigorous phantom studies, the co-registration accuracy of ULM and PAI was assessed to be 20 µm ± 11 µm. The calibration and phantom fabrication techniques are described in detail to enable this type of measurement to be reproduced and used to benchmark other systems. Assessment of oxygen saturation in an oxygen saturation-mimicking solution yielded estimation errors below 5%, even when only five optical wavelengths were considered. ULM provided excellent spatial continuity and high-resolution reconstruction of vascular geometries. The reconstructed 3D density maps and flow measurements achieved micrometre-scale precision. Finally, the *in vivo* experiments demonstrated the feasibility of obtaining simultaneous mapping of oxygenation and perfusion. Such capability is of major interest for studying a myriad of pathological changes, for example during tumour progression and therapy. In conclusion, this paper presents a rigorous approach to co-register ULM and PAI while showing that sufficient sensitivity for preclinical imaging can be obtained using a planar matrix array. Although the fundamental directivity artifacts associated with PAI of vascular structures using such a device remain a limitation, the data presented here provide a useful foundation for future research on hybrid systems and innovative signal processing approaches liable to overcome this limitation.

## Acknowledgements

This work was partly funded by France Life Imaging (grant ANR-11-INBS-0006) and received financial support from the 2021–2030 Cancer Control Strategy through funds administered by Inserm. L.D. acknowledges funding from the Interfaces Pour le Vivant doctoral program at Sorbonne Université.

*In vivo* imaging was performed at the Life Imaging Facility of Université Paris Cité (Plateforme Imageries du Vivant—PIV), supported by France Life Imaging (grant ANR-11-INBS-0006), Région Île-de-France, SIRIC CARPEM (grant INCa-DGOS-Inserm-ITMO Cancer_18006), the IBISA consortium, and financial support from ITMO Cancer of Aviesan within the framework of the 2021–2030 Cancer Control Strategy, through funds administered by Inserm.

This work was partly funded by the Cancer Research for Personalized Medicine – CARPEM project (Site de Recherche Intégré sur le Cancer, SIRIC), the Plan Cancer Physicancer (grant number C16025KS), and the Région Île-de-France SESAME I4M program (grant no. 00000120). B.T. gratefully acknowledges support from the European Union through the European Innovation Council Pathfinder Open Programme (RETIMAGER, grant 101099096).

## Conflict of interest

O.C. is a co-inventor on several patents related to Ultrasound Localization Microscopy and a co-founder of ResolveStroke.